\documentclass[aps,prx,onecolumn,superscriptaddress,nofootinbib,notitlepage,longbibliography]{revtex4-1}
\usepackage{amsbsy}
\usepackage{amsfonts}
\usepackage{amsmath}
\usepackage{amstext}
\usepackage{amsthm}
\usepackage{amssymb}
\usepackage{bm}
\usepackage{color}
\usepackage{comment}
\usepackage{booktabs}
\usepackage{datetime}
\usepackage{dsfont}
\usepackage{dcolumn}
\usepackage{esint}
\usepackage{extarrows}
\usepackage{graphicx}
\usepackage[colorlinks,citecolor=blue]{hyperref} \usepackage{cleveref}
\usepackage{longtable,booktabs}
\usepackage{mathrsfs}
\usepackage{multirow}
\usepackage[super]{nth}
\usepackage{subfigure}
\usepackage{tensor}
\usepackage{url}
\usepackage{xfrac}

\hypersetup{colorlinks=true, linkcolor=blue, filecolor=magenta, urlcolor=blue,}

\begin{document}
\graphicspath{{figures/}}

\title{Mottness, Phase String, and High-$T_c$ Superconductivity}
\thanks{This brief review is an extension of a contribution to the volume of \emph{Festschrift for the Yang Centenary} in celebration of Prof. C. N. Yang's 100th birthday. }%

\author{Jing-Yu Zhao} \affiliation{Institute for Advanced Study, Tsinghua University, Beijing 100084, China}
\author{Zheng-Yu Weng} \affiliation{Institute for Advanced Study, Tsinghua University, Beijing 100084, China}

\date{\today}

\begin{abstract}
It is a great discovery in physics of the twentieth century that the elementary particles in Nature are dictated by gauge forces, characterized by a nonintegrable phase factor \cite{Yang1974} that an elementary particle of charge $q$ acquires from A to B points:
\begin{equation}
P \exp \left( i \frac q {\hbar c}\int_A^B A_{\mu}dx^{\mu}\right) 
\label{nonit}
\end{equation}
where $A_{\mu}$ is the gauge potential and $P$ stands for path ordering. In a many-body system of strongly correlated electrons, if the so-called Mott gap is opened up by interaction, the corresponding Hilbert space will be fundamentally changed. A novel nonintegrable phase factor known as phase-string will appear and replace the conventional Fermi statistics to dictate the low-lying physics. Protected by the Mott gap, which is clearly identified in the high-$T_c$ cuprate with a magnitude $> 1.5 \ \mathrm{eV}$, such a singular phase factor can enforce a fractionalization of the electrons, leading to a dual world of exotic elementary particles with a topological gauge structure. A non-Fermi-liquid ``parent'' state will emerge, in which the gapless Landau quasiparticle is only partially robust around the so-called Fermi arc regions, while the main dynamics are dominated by two types of gapped spinons. Antiferromagnetism, superconductivity, and a Fermi liquid with full Fermi surface can be regarded as the low-temperature instabilities of this new parent state. Both numerics and experiments provide direct evidence for such an emergent physics of the Mottness, which lies in the core of a high-$T_c$ superconducting mechanism.
\end{abstract}

\maketitle

\tableofcontents

\section{Introduction}

\begin{figure}[t]
\begin{center}
\includegraphics[width=1.0\textwidth]{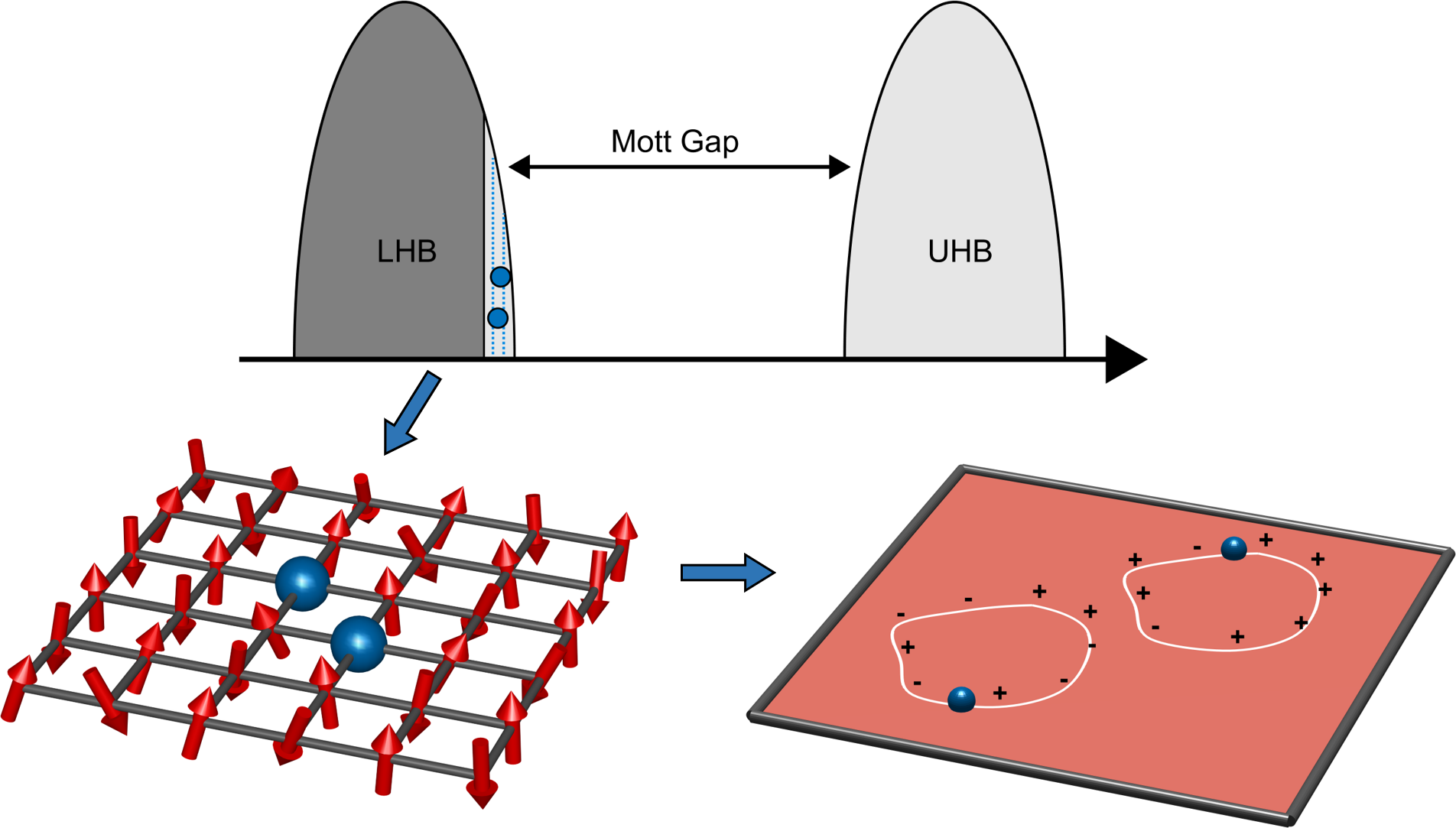}
\end{center}
\caption{The Mott gap due to a local interaction splits the single band of the electrons into the upper Hubbard band (UHB) and lower Hubbard band (LHB). A nonintegrable phase factor as a product of signs (cf. Fig. \ref{psfig}), will be generally picked up by a dopant moving in the spin background, which is composed of the electrons in the filled LHB. The novel quantum interference of such a phase-string effect will replace the conventional Fermi statistics in the Landau paradigm as a new organizing principle, which dictates the essential physics of the doped Mott insulator. }
\label{Mottgap}
\end{figure}

As a revolutionary idea in the twentieth century, the gauge symmetry principle dictates the fundamental interactions between the elementary particles, which is completely characterized \cite{Yang1974} by a nonintegrable phase factor [Eq. (\ref{nonit})]. In stark contrast to the \emph{real} elementary particles with gauge interactions in high-energy physics, we will point out in this article that a novel world of elementary particles with gauge interactions can also \emph{emerge} in a condensed matter system of strongly correlated electrons purely based on the non-relativistic quantum mechanics with long-range entanglement. In the low-energy regime of such a many-body system, the characterization of electronic single-particle excitations can be qualitatively different from the conventional Landau's quasiparticles in the so-called Landau paradigm \cite{AGD1963,Pines1965} which has been a pillar stone of the modern condensed matter physics.  In other words, the most  elementary excitations can become some fractionalized objects other than electron-like. Gauge interactions also arise naturally among these fractional particles.  Such elementary particles with a novel gauge structure constitute the precise mathematical description for a class of many-body systems known as the doped Mott insulator as illustrated in Fig. \ref{Mottgap}, in which a nonintegrable phase factor similar to Eq. (\ref{nonit}) emerges as a new organizing principle. It is protected by the Mott gap, which is caused by the local interaction to split the single-band of the electron into the so-called lower and upper Hubbard bands, respectively \cite{LNW2006}. 

At half-filling, the filled lower Hubbard band is charge neutral, which is known as the Mott insulator. Its spin degrees of freedom remain unfrozen, which can be adiabatically connected to the Heisenberg antiferromagnet in the large Mott gap limit. The doped Mott insulator will be quite different from a semiconductor as the dopants - doped holes in the lower Hubbard band or doped electrons in the upper Hubbard band - can strongly interact with the many-body spin background of the Mott insulator. Thirty five years ago, P. W. Anderson seminally proposed \cite{pwa,pwa1} that the doped Mott insulator lies in the core physics of the high-$T_c$ cuprate. The presence of a Mott gap $> 1.5 \ \mathrm {eV}$ in the cuprate superconductors has been well established experimentally today \cite{Wang2012}.

The state-of-the-art experiments in condensed matter physics have provided a rich variety of means to investigate the realistic materials like the cuprate. In principle, it is an overdetermined problem as so many different experimental probes are available to  ``interrogate'' the same compounds from spin, charge, and electron channels, from transport to spectroscopies, from low energy to high energy, from low doping to high doping \cite{Zaanen2015}.  This is presumably an advantage of condensed matter physics over high energy physics where extremely high energy accelerators are needed in order to smash atoms or nuclei or so into more elementary constituents. Nevertheless, a great experimental complexity so far observed in the cuprate phase has actually contributed to more confusions instead of clarifying the fundamental nature of the high-$T_c$ problem based on the conventional wisdoms.

A full understanding of the Mott physics (dubbed ``Mottness'') is thus essential to provide the correct key ideas and theoretical framework in solving the high-$T_c$ problem. In particular, the aforementioned fractionalized particles and gauge structure of the Mottness identified in a rather simple and exotic \emph{duality world} should be properly translated back to the extremely complex real world of the electrons in order to explain the experiment. In this article, we will briefly review a recent systematic effort in examining the non-perturbative, strongly correlated nature of the Mottness mainly from the concept of phase string \cite{Weng2007,Zaanen2009,Weng2011r}.  Alternative approaches to the Mottness have been reviewed elsewhere \cite{LNW2006,pwa2,gros07}. For a gapped bulk system without gapless modes, the general problem of quantum long-range entanglement has been also systematically investigated and classified based on the concept of ``topological order" \cite{Wen2019}.   

Since the Mottness is based on a lattice model, e.g., a large-$U$ Hubbard model or the $t$-$J$ model or the like, which mainly involves the local hopping and strong interactions of the electrons, exact numerical algorithms will be available as another powerful machinery in the study. Of course, to truly access the long-wavelength, low-energy physics, large-scale numerics may be required which is still an NP hard problem due to an exponentially large Hilbert space with the increase of the sample size. However, with a relatively large Mott gap as revealed in experiment, the unique Mott properties can emerge already at a length scale comparable to the lattice constant, which is well accessible by a finite-size numerical simulation. In the following, we will first present a precise organizing principle of the Mottness based on the $t$-$J$ and Hubbard models and then show that nontrivial physical consequences with non-perturbative nature can indeed occur at any finite-size lattices, which may be thus properly examined numerically.

\section{Organizing Principle of the Mottness}

A doped Mott insulator has been widely recognized as a strongly correlated system where perturbative approaches fail. How to establish a correct description for the Mott physics has been one of the most challenging and difficulty issues of the condensed matter physics in the past three decades. In the following, two organizing principles for the Mottness will be presented, which underlie the basic understandings of the nature of strong correlations and high-$T_c$ superconductivity presented in this article.     

Fermi statistics of the electrons and the corresponding sign structure of quantum many-body wave functions, which enforces the Pauli exclusion principle, are the foundation of Landau's Fermi liquid description in the traditional condensed matter physics.  However, such a basic law will be fundamentally changed once the Mott gap is opened up by interaction, leading to a new and novel sign structure that is to be elaborated as follows.

\subsection{Sign structure of the Mottness: Phase-string} \label{2A}

 \begin{figure}[h]
\begin{center}
\includegraphics[width=0.8\textwidth]{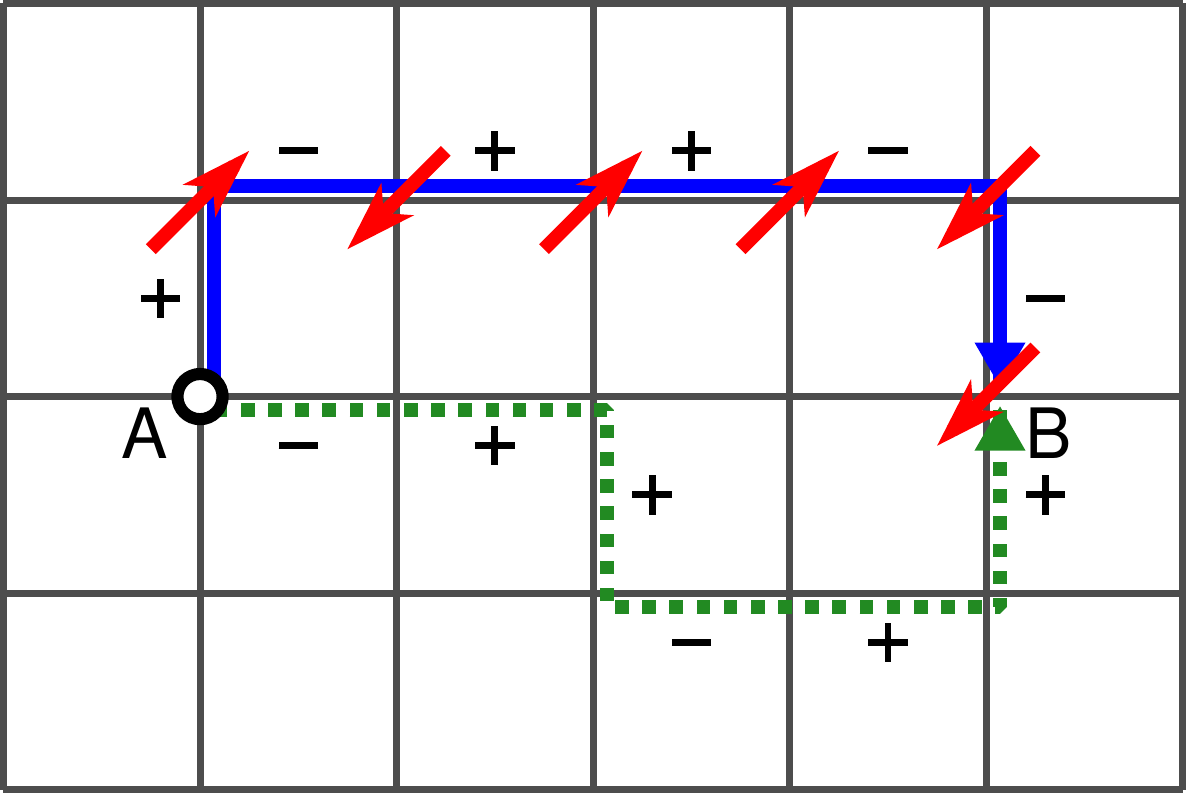}
\end{center}
\caption{Phase strings as products of signs given in Eq. (\ref{ps}), which are picked up by a hole moving from A to B from distinct paths. Here the sign $\pm $ depends on the backflow spin $\sigma=\pm 1$ (as explicitly illustrated for the blue-colored path) exchanged with the hole at each step of hopping. }
\label{psfig}
\end{figure}

Without loss of generality, let us first focus on the $t$-$J$ model, which is equivalent to the Hubbard model at $U/t\gg 1$ limit on a bipartite lattice [$U$ denotes the onsite Coulomb repulsion and $t$ the nearest neighbor (NN) hopping integral]. An exact theorem, which holds true for arbitrary dimensions, electron concentration, and temperature, has been established \cite{Weng1996,WWZ2008}. When a doped charge carrier moves from A to B in Fig. \ref{psfig}, it will always pick up a nontrivial phase factor:
\begin{equation}
\tau_{c}\equiv (+1)\times (-1)\times (-1)\times \cdot\cdot\cdot
\label{ps}
\end{equation}
as a product of sign $\pm$ that faithfully records the backflow spin $\sigma=\pm 1$ (in the $S^z$-quantization direction) at each step of hopping along a path $c$ connecting A and B (two distinct paths are shown in Fig. \ref{psfig}). Such a phase factor is known as the phase-string,  which, as a path-dependent (nonintegrable) phase factor similar to Eq. (\ref{nonit}), will precisely appears in the single-particle propagator, the ground state energy expression, and the partition function \cite{Weng1996,WWZ2008}.

For example, the partition function can be generally expressed by \cite{WWZ2008}  
\begin{equation}
Z_{\mathrm {t-J}}=\sum_{\{c\}}\tau_{c }\tau^{hh}_c {W}[\{c\}]
\label{Z}
\end{equation}
where $\{c\}$ represents the closed loops of all the dopants and spins, with a positive weight ${W}[\{c\}]\geq 0$
depending on the coupling constants, temperature, and dopant concentration \cite{WWZ2008}.  The conventional Fermi statistical sign structure, $(-1)^{N_{\mathrm {ex}}(\{c\})}$, with $N_{\mathrm {ex}}$ denoting the number of exchanges of the whole \emph{electrons} as identical particles \cite{Wu1984}, is significantly reduced to $\tau_{c }\tau^{hh}_c$ in Eq. (\ref{Z}), in which the residual Fermi statistical signs are for \emph{dopants} only as given by
\begin{equation}
\tau^{hh}_{c}\equiv (-1)^{N^h_h(\{c\})},
\label{hh}
\end{equation}
where $N^h_h(\{c\})$ merely counts the total exchanges between the \emph{doped} charge carriers.

The phase-string sign structure $\tau_{c}$ can be also precisely determined in a finite-$U$ Hubbard model \cite{Zhang2014} associated with doped holes and doped electrons, respectively. As illustrated by Fig. \ref{Mottgap}, the opening of a Mott gap due to the local repulsion between the electrons at half filling will lead to a Mott insulator with the lower Hubbard band filled up by the electrons. When such a Mott insulator is doped by holes or electrons in the lower or upper Hubbard band, the dopants will generally pick up a Berry-phase-like phase string as given in Eq. (\ref{ps}). In other words, each dopant will induce a many-body response from the background in the form of the phase-string effect, which makes a doped Mott insulator fundamentally distinct from the picture of a doped semiconductor where only the Fermi sign structure Eq. (\ref{hh}) remains without $\tau_{c}$. Finally, the same sign structure $\tau_{c}$ is still present in a geometrically frustrated model like the $t$-$t'$-$J$ model [with $t'$ the next-nearest-neighbor (NNN) hopping integral] or a triangular Hubbard model with merely an additional geometric (non-dynamic) Berry phase appearing in Eq. (\ref{Z}), whose proofs will be given in detail elsewhere. Therefore, in general, the phase-string sign structure is a precise property of a doped Mott insulator with spin rotational symmetry, which is protected by the Mott gap.

It has been previously conjectured by Anderson \cite{Anderson1990} that an unrenormalizable Fermi-surface phase shift will be induced by each dopant in a doped Mott insulator, which causes an ``orthogonality catastrophe''  \cite{Anderson1967a,Anderson1967b} and leads to a non-Fermi-liquid behavior of the many-body ground state. The phase-string sign structure identified rigorously in the $t$-$J$ and Hubbard models corresponds precisely to such a many-body phase shift. But $\tau_{c}$ in Eqs. (\ref{ps}) and (\ref{Z}) is more singular as a nonintegrable phase factor, which is not only prominent near the Fermi surface as originally speculated by Anderson \cite{Anderson1990}. As a matter of fact, a spin flip can lead to a total sign change in $\tau_{c}$. As such, the whole spin background of the Mott insulator becomes effectively long-range entangled with the dopants in a sufficiently long-time, low-energy limit. Thus the phase-string factor $\tau_{c}$ implies a very non-perturbative nature of the Mottness that calls for a new methodology drastically different from the conventional many-body approaches \cite{AGD1963,Pines1965}. 

\subsection{Mutual duality: An emergent topological gauge structure} \label{2B}

The phase-string sign structure outlined above will replace the conventional Fermi sign structure as a new organizing principle of the Mottness. To understand its nontrivial physical implications, in the following a mutual-duality transformation is introduced \cite{Weng1997,Weng1998,Weng2011}, which can effectively regulate and smoothen the singular hopping term, and reveal a hidden equivalence of the phase-string with a topological gauge structure. 

Make a unitary transformation as follows 
\begin{equation}
|\Psi\rangle =e^{i\hat{\Theta}} |{\Phi}\rangle ,
\label{unitary}
\end{equation}
which is defined by \cite{Weng1997,Weng1998,Weng2011}     
\begin{equation}
e^{i\hat{\Theta}} \equiv e^{-i\sum_i n_i^h \hat{\Omega}_{i}} ,
\label{unitary2}
\end{equation}
where $n_i^h$ defines the number operator of the dopants (doped holes, without loss of generality) at site $i$. Each doped hole introduces a nonlocal phase shift $\hat{\Omega}_{i}$, which is defined by
\begin{equation}
\hat{\Omega}_{i}\equiv \frac{1}{2} \sum_{l\neq i}\theta _{i}(l)\left( 1-\sum_{\sigma }\sigma
n_{l\sigma }^b \right)  \label{phif}
\end{equation}%
with $\theta _{i}(l)=\mathrm{Im}\ln $ $(z_{i}-z_{l})$ (where $z_i$ is the complex coordinate of the lattice site $i$) and $n_{l\sigma }^b$ denotes the number of the background spin-$\sigma$ at site $l$ in two-dimensional (2D) square lattice.  
Then, accordingly the $t$-$J$ Hamiltonian can be transformed into a new Hamiltonian acting on the dual space $ |{\Phi}\rangle $: 
\begin{equation}
H_{\mathrm { t-J}} \rightarrow H_{\mathrm {\sigma t-J}} [A^s-\phi^0; A^h],
\label{tjstj}
\end{equation}
where the three NN link variables, $A_{ij}^s$, $\phi_{ij}^0$, $A_{ij}^h$, are defined by
\begin{eqnarray}
A_{ij}^s &=& \frac 1 2\sum_{l\neq i,j} \left[\theta_i(l)-\theta_j(l)\right] (n^b_{l\uparrow}-n^b_{l\downarrow}), \label{MCS01}\\
\phi^0_{ij} &=&\frac 1 2 \sum_{l\neq i,j}  \left[\theta_i(l)-\theta_j(l)\right] , \label{MCS02}\\
A_{ij}^h &=&\frac 1 2 \sum_{l\neq i,j} \left[\theta_i(l)-\theta_j(l)\right] n^h_l .
\label{MCS03}
\end{eqnarray}

It is straightforward to verify, by setting $A^s=\phi^0=A^h=0 $, that the so-called $\sigma\cdot $$t$-$J$ Hamiltonian, $H_{\mathrm {\sigma} t-J} [0;0]$, on the right-hand-side of Eq. (\ref{tjstj}) will give rise to the following partition function \cite{Zhu2013}
\begin{equation}
Z_{\mathrm {\sigma t-J}}=\sum_{\{c\}}\tau^{hh}_c {W}[\{c\}]
\label{Zs}
\end{equation}
where the phase-string factor is precisely ``switched off'', while the dopants' fermion sign structure $\tau^{hh}_c$ and the weight functional ${W}[\{c\}]$ remain the same as in the $t$-$J$ case [Eq. (\ref{Z})]. 

Therefore, we arrive at the second exact theorem based on the $t$-$J$ model. It states that the phase-string sign structure is precisely equivalent to three NN link variables, $A_{ij}^s$, $\phi_{ij}^0$, $A_{ij}^h$, namely,
\begin{equation}
\tau_{c} \leftrightarrow  A^s, \phi^0, A^h
\label{}
\end{equation}
in a duality world  (note that it only applies to the low-dimensions like 2D, in contrast to the exact phase-string sign structure which holds true for arbitrary dimensions).  

Note that the duality transformation (\ref{unitary2}) is not unique. The simplest redundancy may be seen by changing $\theta _{i}(l)\rightarrow \theta _{i}(l)+\phi$ with an arbitrary constant $\phi$, which does not change $A_{ij}^s$, $\phi_{ij}^0$, $A_{ij}^h$ or the dual Hamiltonian. In general, it is a gauge redundancy and these three link variables are gauge fields subject to $U(1)$ symmetries, respectively, in the transformed Hamiltonian and can only be determined up to their gauge fluxes given as follows. Namely, their oriented-closed-loop summations satisfy the following gauge-invariant topological constraints \cite{Weng1997,Weng1998,Weng2011}
\begin{eqnarray}
\sum_c A_{ij}^s &=&\pm \pi \sum_{l\in c} (n^b_{\uparrow}-n^b_{\downarrow}), \label{MCS1}\\
\sum_c \phi^0_{ij} &=&\pm \pi \times \left({\mathrm {\# \ of \ plaquette \ enclosed \ by \ c}}\right), \label{MCS2}\\
\sum_c A_{ij}^h &=&\pm \pi \sum_{l\in c} n^h_l ~.
\label{MCS3}
\end{eqnarray}
Here the field strength of $A^s$ is generated by the polarization (along the $S^z$-quantization axis) of spins in the background, the strength of $A^h$ is generated by the local density of dopants, while the strength of $\phi^0$ is a constant $\pi$ per plaquette. $A^s$ and $A^h$ are known as the mutual Chern-Simons gauge fields \cite{Weng1997,Weng1998,Weng2011}, which imply a mutual semionic statistics \cite{Wilczek1990} between the dopants and background spins as introduced by the phase-string effect.

Furthermore, in such a transformed dual world, the low-lying physics becomes substantially smoothened. For instance, in a finite doping regime where the antiferromagnetic correlation between the background spins presumably becomes short-ranged, and the dopants have a uniform distribution on lattice, then, one has  $A_{ij}^s \simeq 0$ and $A_{ij}^h$ describing a uniform flux of $\delta \pi$ per plaquette ($\delta$ is the doping concentration) to the leading order of approximation (up to a pure $U(1)$ gauge). Recall that the $\sigma\cdot $$t$-$J$ Hamiltonian, $H_{\mathrm{\sigma t-J}} [0;0]$, is phase-string free, which actually describes a conventional Fermi liquid system for the dopants as to be shown in the exact numerics below. Thus, the full duality Hamiltonian $H_{\mathrm {\sigma t-J}} [A^s-\phi^0;A^h]$ of the $t$-$J$ model will lead to a more conventional-looking ground state to be discussed later. In other words, even though the phase-string sign structure in Eq. (\ref{ps}) is very singular in the original $t$-$J$ model, its many-body quantum interference can lead to a dual description in which the constituents as fractionalized objects, with the gauge charges of  $A^s$, $\phi^0$, $A^h$, respectively, can behave much more coherent and consequently perturbatively treatable. 

\subsection{Numerical verification of the phase-string effect} \label{2C}

Based on the above two \emph{exact} theorems, one may critically compare the ground state results between the $t$-$J$ and $\sigma\cdot $$t$-$J$ models. Two models differ only by the presence and absence of the phase-string factor $\tau_{c}$ in Eqs. (\ref{Z}) and (\ref{Zs}), respectively. Thus the pure and singular effect of the sign structure can be accurately singled out and verified by finite-size numerics.  

Since the theorem on the phase-string sign structure holds true for any lattices, one can study a 2D sample of size $N=L\times L$, or a ladder of size $N=L \times 1$ (chain), $N=L \times 2$ (two-leg), $N=L \times 3$ (three-leg), $N=L \times 4$ (four-leg), .... Due to the particle-hole symmetry in the Hubbard model, one may only focus on the hole doped side, with single hole, two holes, ... , up to a finite doping concentrations of dopants. 
  
\subsubsection{Single-hole case}

It is noted that at half-filling, the $t$-$J$ and $\sigma\cdot $$t$-$J$ models reduce to the same Heisenberg model, whose ground state is spin-singlet with long-range antiferromagnetic correlations in a large sample \cite{Marshall}. Single hole injected into such an antiferromagnetic spin background will induce the phase-string sign structure in the $t$-$J$ model but none in the $\sigma\cdot $$t$-$J$ model. Therefore, the single hole problem is the simplest case to demonstrate the phase-string effect by simply comparing the ground state properties between the two models, which can be calculated by exact diagonalization (ED) and density matrix renormalization group (DMRG) numerical methods. In the following we briefly outline some important results.

In 2D with $C_4$ rotational symmetry and open boundary condition (OBC), the single-hole ground state has been shown to have a fourfold degeneracy in a finite sample. In addition to a simple double degeneracy of $S^z=\pm 1/2$ (since one spin-1/2 is removed from the spin-singlet ground state), a novel angular momentum of $L_z=\pm 1$ has been revealed in ED ($L=4$) and DMRG ($L=6$, $8$) results \cite{Zheng2018}. Furthermore, corresponding to $L_z=\pm 1$, a spin current which conserves $S^z=\pm 1/2$ is found in the background. On the other hand, for a 2D sample with periodic boundary condition (PBC), the novel double degeneracy of $L_z=\pm 1$ will be replaced by a fourfold degeneracy characterized by four total momenta at $(\pm\pi/2,\pm\pi/2)$ in the ground states. Again the spin currents are found in the background as the backflow associated with the doped hole to generate a finite momentum in the ground state \cite{Zheng2018}.  

Physically the novel quantum $L_z$ and the spin currents in the ground state are a direct manifestation of the phase-string effect in Eq. (\ref{ps}), which means that accompanying each step of hole hopping, the backflow must be spin-dependent to explicitly exhibit a spin current if $S^z\neq 0$ as shown in the exact numerics. Indeed, once one artificially switches off the phase string in the $t$-$J$ model to result in the $\sigma\cdot $$t$-$J$ model, the numerical calculations have shown \cite{Zheng2018} that the novel quantum $L_z$ and the spin currents all disappear with only the trivial double degeneracy due to $S^z=\pm 1/2$ left in the new ground state under OBC. Furthermore, under PBC, the total momentum reduces to the trivial symmetric point at $(0,0)$ or $(\pi,\pi) $ with restoring the translational symmetry without the presence of spin currents in the background.

Similar comparative studies of the single hole in the one-dimensional chain \cite{Zhu2016,Zheng2018r}, two-leg, three-leg, and four-leg ladders, etc. \cite{Zhu2013,Zhu2015,Zhu2018,Sun2019}, as well as in the Hubbard model \cite{He2016,Ho2016,Tohyama2021}, all have shown that in the $t$-$J$ (Hubbard) case the doped hole always generates a nontrivial backflow spin current with new quantum number, which involves the many-body spin excitations from the background, while in the $\sigma\cdot $$t$-$J$ ($\sigma$-Hubbard) case, the doped hole simply behaves as a Bloch wave (i.e., a Landau-like quasiparticle with translational symmetry), which is essentially decoupled from the half-filled spin background. The presence/absence of the spin current is thus an explicit manifestation of the presence/absence of the phase-string effect in these models. 

\subsubsection{Two-hole case}

The above numerical studies have unequivocally demonstrated that the phase-string sign structure of Eq. (\ref{ps}) has played a crucial role to turn a Landau-like quasiparticle (in the $\sigma\cdot $$t$-$J$ model) into a ``twisted'' quasiparticle with novel quantum number and an explicit spin current formed by its spin-1/2 partner in a finite-size sample with different geometric shapes of the lattice. 

By a sharp contrast, the aforementioned novel degeneracy and underlying spin current disappear when two holes are injected into the antiferromagnetic background, leading to a strong pairing between the holes. The 2D two-hole ground state is characterized by spin singlet, angular momentum $L_z= 2\ \mathrm{mod \ 4}$, which is nondegenerate with the $C_4$ rotational symmetry of the square sample under OBC \cite{Zheng2018}. Two-leg and four-leg ladders all show \cite{Zhu2014} strong binding energies for the two holes in the nondegenerate ground state. In particular, in the two-leg ladder, one may continuously tune the rung (interchain) hopping integral into zero, where the pairing state remains and the phase-string can be analytically sorted out. Then it can be precisely shown \cite{Zhu2018pairing} that the pairing energy is gained by eliminating the frustration effect due to the phase-string on each individual hole. Simultaneously the spin current is eliminated too there by pairing.

Such a new pairing mechanism due to the phase-string effect can be further directly seen from the fact that when the phase-string is turned off in the $\sigma\cdot $$t$-$J$ model, the pairing vanishes too and the two-hole ground state simply reduces to that of two Landau quasiparticles decoupled from the spin background. Note that in the two-leg and four-leg cases, the undoped spin backgrounds are in short-range antiferromagnetic or the RVB states, and previously Anderson conjectured \cite{pwa} the so-called RVB pairing mechanism for superconductivity based on the $t$-$J$ model. However, the above numerical simulations clearly indicate that the RVB is \emph{not} dominantly responsible for the pairing of dopants, as shown in the $\sigma\cdot $$t$-$J$. It is the phase-string induced by the motion of doped holes in the $t$-$J$ model that forces a strong binding between two holes in order to reduce its frustration effect. 

\subsubsection{Finite-doping case}

As shown by numerical calculations, the strong pairing identified in the two-hole case can persist continuously into a finite-doping regime. Here the pairing mechanism can be well established in a finite-size study as the pairing size is generally small. It thus indicates that either a superconductivity or a phase separation may naturally arise in a finite-doped Mott insulator in the thermodynamic limit. 

By DMRG calculation, two-leg $t$-$J$ ladder has been investigated, say, at doping concentration $\delta=1/12$, $1/8$, etc., where quasi-long-range pair-pair correlation has been found at large $L$. The ground state as a quasi-1D superconducting state known as the Luther-Emery (LE) state has been confirmed \cite{Jiang2020}. Various pseudogap-like phenomena have been also found when the spin background is artificially tuned into a long-range antiferromagnetic order \cite{Sun2020} or partially polarized by Zeeman fields, resulting in a charge-density-wave or pair-density-wave order with the pairing becoming subleading. For the four-leg ladder, similar complex phase diagram has been also found \cite{Jiang2018}, in which an NNN hopping integral $t'$ may be further needed to stabilize the LE phase from the stripe and phase separation instabilities \cite{Jiang2019}. Rich properties have been also reported \cite{White2022} in the literature for the six-leg case, which is even closer to the 2D limit.

Then it is important to show that once the phase-string sign structure is turned off, the above mentioned pairing, stripe or pair density wave order in the complex phase diagram of the $t$-$J$ ladders have been found all gone, with the ground states simply behaving like a Fermi gas formed by the dopants on top of a decoupled spin background in the $\sigma\cdot $$t$-$J$ ladders. In other words, it proves \cite{Jiang2020,ZhangJX2022} that the pairing and the various charge ordering states in the $t$-$J$ (Hubbard) model can all be attributed to the phase-string effect, which induces the essential long-range mutual entanglement between the dopants (holes) and spin background that has been turned off in the $\sigma\cdot $$t$-$J$ ($\sigma$-Hubbard) model. 

Therefore, the exact numerics have indeed accurately demonstrated that the phase-string sign structure plays an essential role in determining the unconventional long-range and short-range physics of the doped Mott insulator, whereas, as soon as the phase-string is turned off, the anomalies in the Mottness disappear to restore a Fermi liquid behavior similar to a doped semiconductor. 

\section{Parent state at finite doping:  A non-Landau paradigm}

In the Mott world, the electrons will be organized according to an emergent law of mutual statistics, i.e., the phase-string sign structure outlined in Sec. \ref{2A}. Without such a novel sign structure, the exact numerics have indicated that the ground state would reduce to a trivial Fermi-gas state composed of the fermionic dopants, which is effectively separated from the spin background of the filled lower Hubbard band, similar to a doped semiconductor. By a sharp contrast, a rich phase diagram including a strong pairing between the dopants will arise once the phase-string sign structure is restored, based on the comparative numerical studies of the $t$-$J$ and $\sigma\cdot $$t$-$J$ models (cf. Sec. \ref{2C}). 

According to the second theorem presented in Sec. \ref{2B}, the duality transformation of Eq. (\ref{tjstj}) incorporates the most singular effect of the phase-string in the mutual swap of a dopant with the backflow spin during the hopping. Then the residual effect of the phase-string becomes smoothened after the transformation, which is precisely described by a novel topological gauge structure. To leading order of approximation, assuming a uniform distribution of charge at finite doping, a \emph{parent ground state} of the $t$-$J$ model has been constructed as follows \cite{Weng2011,Ma2013}  
\begin{equation}
|\Psi _{\mathrm{G}}\rangle=e^{i\hat{\Theta}}|{\Phi} _{\mathrm{G}}\rangle ~,
\label{gs}
\end{equation}
where
\begin{equation}
|{\Phi} _{\mathrm{G}}\rangle \propto \exp \left(\sum_{ij}g_{ij}\hat{c}_{i\uparrow }\hat{c}_{j\downarrow } \right) |\text{b-RVB}
\rangle ~ . \label{frac1}
\end{equation}
The ground state ansatz in Eqs. (\ref{gs}) and (\ref{frac1}) is fundamentally distinct from the original proposal of a \emph{single-component} RVB state by Anderson \cite{pwa}. Here $|{\Phi} _{\mathrm{G}}\rangle $ is composed of a \emph{two-component} RVB structure: $|\text{b-RVB} \rangle$ denotes a \emph{neutral} RVB spin background, and a BCS-like component is created on the vacuum state $|\text{b-RVB} \rangle$ in terms of a Cooper-like pair of doped holes, $ \sum_{ij}g_{ij}\hat{c}_{i\uparrow }\hat{c}_{j\downarrow } $. At half filling, $|\Psi _{\mathrm{G}}\rangle$ reduces to $|\text{b-RVB} \rangle$, and the latter recovers the so-called long-range bosonic RVB state, which was first proposed by Liang, Doucot, Anderson \cite{Liang1988} to accurately describe a long-range antiferromagnetic state of the Heisenberg model. At finite doping,  
the vacuum state $|\text{b-RVB} \rangle$ will evolve into a short-range antiferromagnetic state (a spin liquid state) self-consistently \cite{Weng2011,Ma2013}. Furthermore, in the original electron representation, an intrinsic long-range mutual entanglement between these two components is introduced via the phase-string factor $e^{i\hat{\Theta}}$ in Eq. (\ref{gs}), whose effect will be implemented via the topological gauge fields in the dual representation. 

In the following, let us first examine the ground-state ansatz in the one-hole and two-hole cases, with using the variational Monte Carlo (VMC) calculations in comparison with the exact numerical results.

\subsection{One-hole and two-hole ground states}
\begin{figure}[t]
\begin{center}
\includegraphics[width=0.8\textwidth]{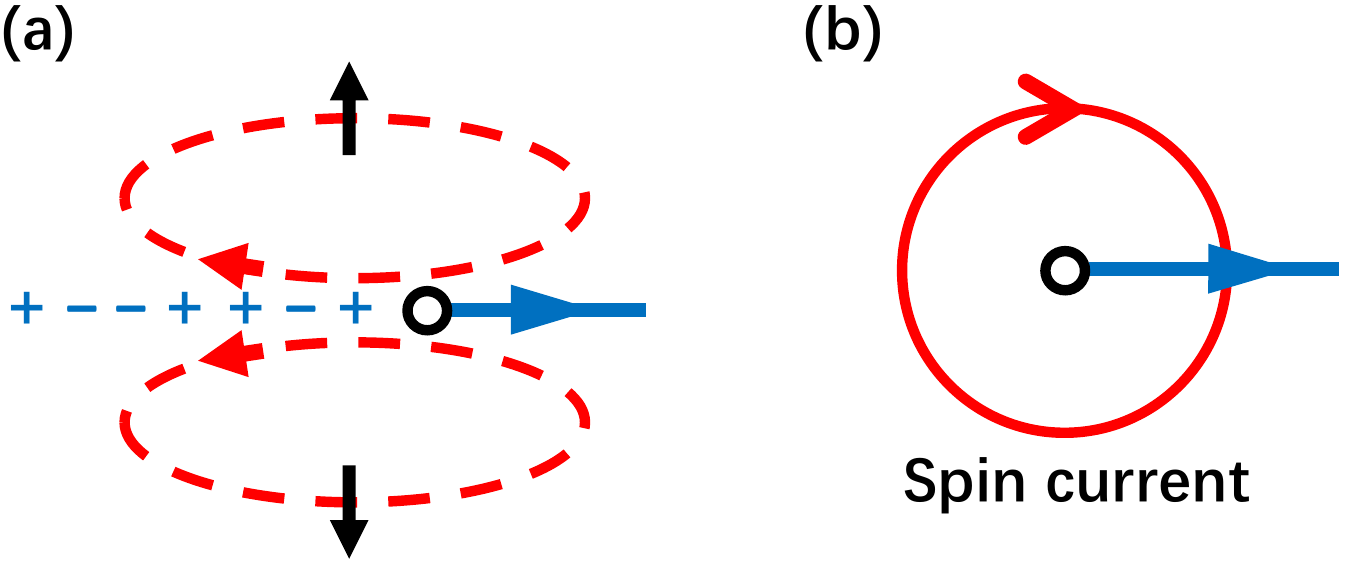}
\end{center}
\caption{The composite structure of a renormalized single hole state in Eq. (\ref{VMC1h}): 
        (a) The motion (blue arrow) of a doped hole (black circle) creates a sequence of $\pm$ signs (phase string) depending on $\uparrow$ and $\downarrow$ of the backflow spins. 
        To avoid cancellation in the hopping integral due to the phase-string effect, the backflow spin currents have to bend over in a roton-like pattern (red-dashed) around the hole; 
        (b) Such a roton configuration may be explicitly manifested by a vortex of spin current around the hole as a \emph{net} effect of an extra $S_z=\pm 1/2$, characterized by an emergent internal angular momentum of $L_z=\pm 1$ in agreement with the exact numerics. The original figure is from Ref. \cite{Zhao2022}.}.    
\label{1hole}
\end{figure}

The simplest nontrivial case of the doped Mott insulator is the single-hole-doped ground state. In the dual space of Eq. (\ref{frac1}) it is simply reduced to 
\begin{equation}
|\Phi_\mathrm{G}\rangle_{\mathrm{1h}}=\sum_{i}\varphi_{\mathrm{h}}\left(i\right)\hat{c}_{i\alpha}|b\text{-RVB}\rangle. \label{VMC1hdual}
\end{equation}
Here 
the single-hole wave function $\varphi_{\mathrm{h}}$ is determined by VMC  \cite{Wang2015,Chen2019}. In contrast, in the original electron representation of Eq. (\ref{gs}), the one-hole ground state is given by
\begin{equation}
|\Psi_\mathrm{G}\rangle_{\mathrm{1h}}=\sum_{i}\varphi_{\mathrm{h}}\left(i\right)\tilde{c}_{i\alpha}|b\text{-RVB}\rangle, \label{VMC1h}
\end{equation}
in which the hole is created by a ``twisted'' hole creation operator
\begin{equation}
\tilde{c}_{i\sigma}\equiv \hat{c}_{i\sigma}e^{-i\hat{\Omega}_{i}}~.
\label{eqn:ctilde}
\end{equation}
Namely the doped-hole wave function is actually $\varphi_{\mathrm{h}}e^{-i\hat{\Omega}_{i}}$, in which the phase-string operator $e^{-i\hat{\Omega}_{i}}$ defined in Eq. (\ref{phif}) plays a crucial role to dress the bare hole by a pattern of spin currents as shown in Fig. \ref{1hole}. It clearly shows that the single-hole ground state cannot be adiabatically connected to a Landau's quasiparticle, whereas its dual description of Eq. (\ref{VMC1hdual})  may be much simpler with a closer resemblance to a Bloch wave. It is noted that here in a finite-size system, $|\text{b-RVB}\rangle$ is simply reduced to the half-filling ground state of the Heisenberg model, although a topological correction may become important in the thermodynamic limit. 

The excellent agreement of the single-hole state in Eq. (\ref{VMC1h}) with finite-size ED and DMRG results outlined in Sec. \ref{2C}, as calculated by the variational Monte Carlo (VMC) method,  including the nontrivial angular momentum $L_z=\pm 1$ in 2D \cite{Chen2019} and the momentum structure in the two-leg ladder case \cite{Wang2015}, clearly demonstrates that the doped hole in the Mottness gets substantially renormalized by the phase-string effect, whose behavior is fundamentally distinct from the celebrated Landau's quasiparticle as a new quasiparticle of the doped Mott insulator. Recently a single-hole-doped case under a perturbation of a weak spin-orbit coupling (SOC) has been also examined \cite{Chen2022}, which has further shown that the twisted hole with the novel angular momentum in Eq. (\ref{eqn:ctilde}) is critical to properly describe the response to the SOC as indicated by an excellent agreement between the VMC and ED.

\begin{figure}[t]
\begin{center}
\includegraphics[width=0.6\textwidth]{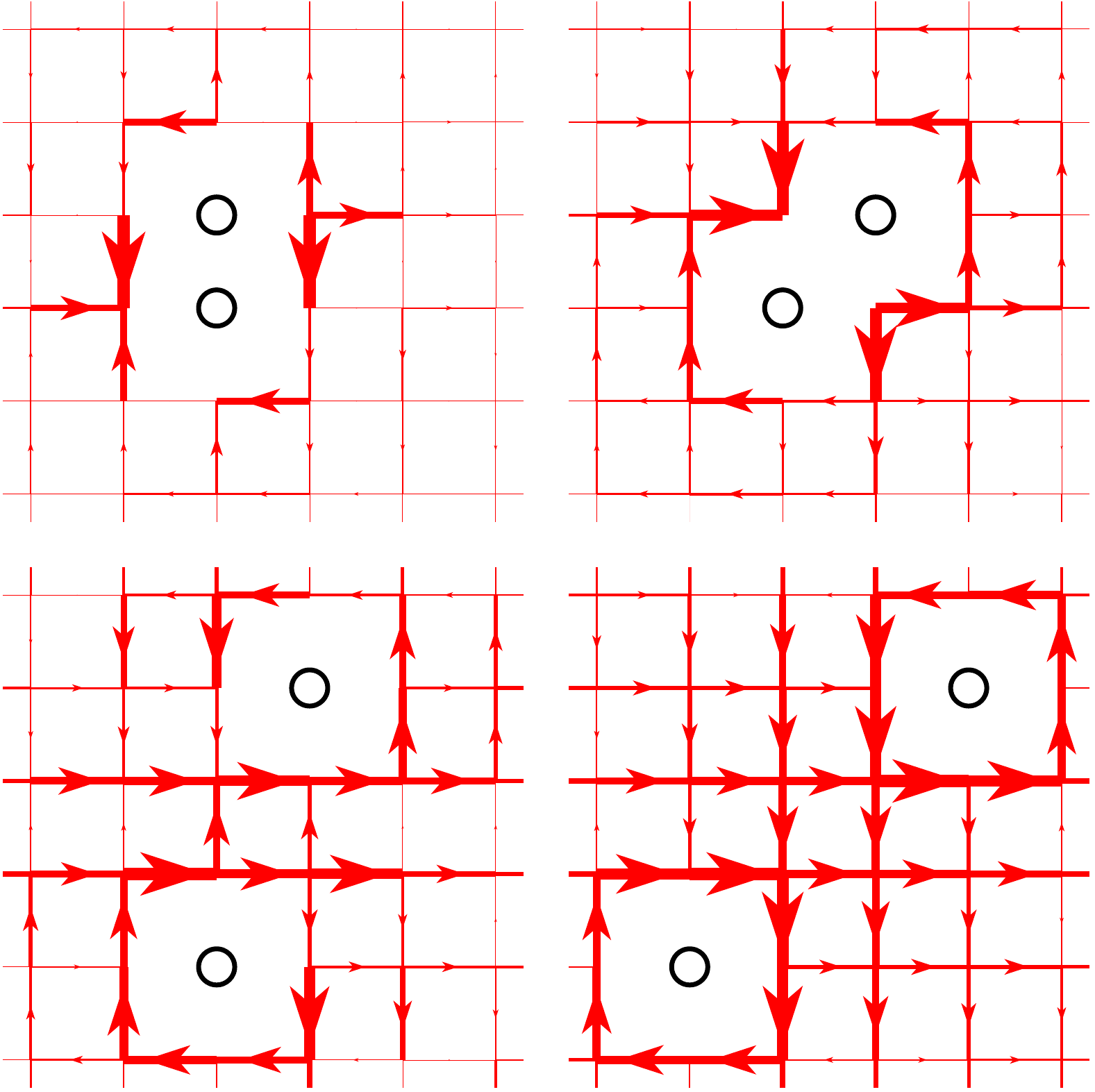}
\end{center}
\caption{A roton-like spin current pattern around two holes \cite{Zhao2022} in the pairing ground state of Eq. (\ref{VMC2h}), which contributes to a strong binding between the two holes with a size $\sim 4a_0\times 4a_0$. }  
\label{2hole}
\end{figure}

Similarly to Eq. (\ref{VMC1h}), the two-hole ground state ansatz based on Eqs. (\ref{gs}) and (\ref{frac1}) may be formally rewritten as \cite{Chen2018,Zhao2022}
\begin{equation}
|\Psi _{\mathrm{G}}\rangle_{\mathrm 2h} =\mathcal{\hat{D}} |\text{b-RVB} \rangle,  \label{VMC2h}
\end{equation}%
with \cite{note1}
\begin{equation}
\mathcal{\hat{D}}\equiv \sum_{ij}g_{ij}\tilde{c}_{i\uparrow }\tilde{c}_{j\downarrow }.  \label{ghat}
\end{equation}%
The VMC calculation in 2D shows \cite{Zhao2022} that the two-hole ground state is nondegenerate with an angular momentum $L_z=2 \mod 4$ under a $C_4$ rotational symmetry, again in agreement with the ED and DMRG results. In particular, a strong binding has been revealed with the pairing mechanism directly connected to the phase-string effect. As shown in Fig. \ref{2hole}, a roton-like spin current pattern originated from the phase-strings has been clearly identified around the two holes. The pairing structure is shown to have a novel dichotomy:  namely, an $s$-wave pairing symmetry is found in $|g_{ij}|$ for the pairing of two twisted quasiparticles $\tilde{c}$, but a $d$-wave symmetry is shown for the overlap of the ground state with the Cooper-pair order parameter. In particular, the strong binding disappears once the phase-string is turned off in Eq. (\ref{eqn:ctilde}) in the VMC calculation, even though the RVB pairing is still present in the spin background $|\text{b-RVB} \rangle$. It implies that the BCS-like pairing of doped holes is predominantly attributed to the phase-string effect \cite{Zhao2022}, instead of the RVB pairing as originally envisaged by Anderson \cite{pwa,pwa1}. Interestingly the pairing size of the two holes is about $4a_0\times 4a_0$ ($a_0$ is the lattice constant), which is much smaller than the spin-spin correlation length in the VMC calculation \cite{Zhao2022}.

\subsection{Finite doping: Fractionalization protected by multiple hidden ODRLOs}

The parent ground state of the doped Mott insulator is basically composed of a two-component structure in Eq. (\ref{frac1}), i.e., the RVB spin background and the BCS-like pairing of the dopants. The RVB pairing amplitude (see below) and the BCS-pairing $g_{ij}$ are presumably smooth functions.  And then the two degrees of freedom are further entangled via $e^{i\hat{\Theta}}$ in Eq. (\ref{gs}). Here $\hat{\Theta} \equiv -\sum_i n_i^h \hat{\Omega}_i $ denotes the total phase-shift induced by doped holes, with $\hat{\Omega}_i $ [Eq. (\ref{phif})] acting on the spin background via $n_{l\sigma }^b $. Note that there is a no-double-occupancy constraint of $n_i^h+\sum_{\sigma}n_{l\sigma }^b=1$ in the earlier formulation \cite{Weng1997}. In the following, it is shown that such a constraint can be replaced by a single-occupancy constraint $\sum_{\sigma}n_{l\sigma }^b=1$ via a further fractionalization \cite{Weng2011}. 

Mathematically, the hole creation operator $\hat{c}_{i\sigma }$ in the dual world of Eq. (\ref{frac1}) may be further reexpressed by \cite{Weng2011} 
\begin{equation}
\hat{c}_{i\sigma }\equiv h_{i}^{\dagger }a_{i\bar{\sigma}}^{\dagger
}\text{ },  \label{decomp}
\end{equation}
with $h^{\dagger}_i$ creating a bosonic holon and $a_{i\bar{\sigma}}^{\dagger}$ a \emph{backflow} spinon which is a fermion. Correspondingly, Eq. (\ref{frac1}) can be rewritten as a three-component ground state
\begin{equation}
|\Phi _{\mathrm{G}}\rangle ={\cal \hat{P}}\left( |\Phi _{h}\rangle \otimes |\Phi
_{a}\rangle \otimes |\Phi _{b}\rangle \right),  \label{gsansatz}
\end{equation}
such that in the duality transformation of Eqs. (\ref{unitary2}) and (\ref{phif}), the holon number operator $n^h_i$ acts on the holon degree of freedom in $|\Phi _{h}\rangle$ and $n_{l\sigma }^{b}$ on the neutral spinon subspace $ |\Phi _{b}\rangle$. 

Here the projection operator $\cal \hat{P}$ in Eq. (\ref{gsansatz}) is defined by
\begin{equation}
{\cal \hat{P}}\equiv \hat{P}_{\mathrm{B}}\hat{P}_{s}\text{ ,}  \label{P}
\end{equation}%
in which $\hat{P}_{s}$ will enforce the single-occupancy constraint $\sum_{\sigma} n_{i\sigma }^{b}=1$ in the spinon state $|\Phi _{b}\rangle $
such that $|\text{b-RVB} \rangle\equiv \hat{P}_{s}|\Phi _{b}\rangle$; and $\hat{%
P}_{\mathrm{B}}$ will further enforce $n_{i\bar{\sigma}}^{a}=n_{i}^{h}n_{i\sigma }^{b}$ such that the $S^z$ spin of the $a$-spinon will compensate that of the $b$-spinon at a hole site.
By applying $\cal \hat{P}$, the physical
Hilbert space is restored in Eq. (\ref{gsansatz}). On the other hand, the $U(1)$ gauge fluctuations associated with the constraint enforced by the projection $\cal \hat{P}$, that is, the feedback effect on the subsystems in Eq. (\ref{gsansatz}) will get suppressed or Higgsed due to the hidden off-diagonal-long-range-orders \cite{Yang-ODLRO} (ODLROs) in the subsystems (see below). The latter provide the rigidity of the spontaneous symmetry breaking to justify this fractionalization scheme self-consistently \cite{Weng2011,Ma2013}. 

In the dual space, the three subsystems in Eq. (\ref{gsansatz}) are given by \cite{Weng2011,Ma2013}
\begin{equation}
|\Phi _{h}\rangle \equiv \sum_{\{l_{h}\}}\varphi
_{h}(l_{1},l_{2},...)h_{l_{1}}^{\dagger }h_{l_{2}}^{\dagger }...|0\rangle
_{h}\ ,  \label{bgs}
\end{equation}%
and
\begin{equation}
|\Phi _{a}\rangle \equiv \exp \left( -\sum_{ij}{g}_{ij}a_{i\uparrow
}^{\dagger }a_{j\downarrow }^{\dagger }\right) |0\rangle _{a}~,  \label{phia-0}
\end{equation}%
as well as%
\begin{equation}
|\Phi _{b}\rangle \equiv \exp \left( \sum_{ij} W_{ij}b_{i\uparrow }^{\dagger
}b_{j\downarrow }^{\dagger }\right) |0\rangle _{b}~.  \label{phirvb}
\end{equation}
Here the holon wavefunction $\varphi _{h}= \text{constant} $ indicates
a Bose-condensed holon state $|\Phi _{h}\rangle $;  $|\Phi _{a}\rangle $ defines a
backflow spinon state with BCS-like ($s$-wave \cite{Ma2013}) pairing
amplitude ${g}_{ij}$; and $ |\Phi _{b}\rangle $ defines a bosonic 
spinon state with an RVB pairing amplitude $W_{ij}$.

\begin{table}[t]
    \centering
    \caption{Fractionalization of the Mottness and the topological gauge fields defined in Eqs. (\ref{MCS1})-(\ref{MCS3}) due to the phase-string sign structure. Here $A^{\mathrm ext}_{ij}$ denotes the external electromagnetic field. }   
 \begin{ruledtabular}
     \begin{tabular}{ccccc} 
    \\
             elementary particle     & holon  & a-spinon & b-spinon & hole  \\  
      
            & $h^{\dagger}_i$ & $a^{\dagger}_{i\sigma} $ & $b^{\dagger}_{i\sigma} $ & $c_{i\sigma} $ \\ \\
            \colrule \\
            
            gauge field & $A^s_{ij}$, \ \ $A^{\mathrm ext}_{ij}$ & $-\phi^0_{ij}$ & $A^h_{ij}$ & $-A^{\mathrm ext}_{ij}$\\ \\
                    \end{tabular}	
\end{ruledtabular}
    \label{tab:particles}
\end{table}

Therefore, the electrons in the parent state are \emph{truly} fractionalized into three more elementary particles, i.e., the bosonic holon with creation operator $h^{\dagger}_i$, the bosonic neutral spinon with creation operator $b_{i\sigma} ^{\dagger}$, and the fermionic backflow spinon with creation operator $a_{i\sigma} ^{\dagger}$. They are coupled to three gauge fields, $A^s_{ij}$, $A^h_{ij}$, and $\phi^0_{ij}$, respectively, which faithfully capture the phase-string sign structure of the Mottness as illustrated in Table \ref{tab:particles}. In particular, the holons and $b$-spinons are also the sources to generate the topological gauge fields, $A^h_{ij}$ and $A^s_{ij}$, as given in Eqs. (\ref{MCS3}) and (\ref{MCS1}), respectively, which lead to a MCS gauge theory description of the low-energy physics \cite{KQW2005,Ye2011}. While both $a$ and $b$ spinons are charge neutral, the holon also carries a positive electric charge and sees the external electromagnetic field $A^{\mathrm {ext}}_{ij}$. 

Finally, in this fractionalization formulation of the parent state, the original bare-hole creation operator can be expressed by
\begin{equation}
{c}_{i\sigma }\equiv h_{i}^{\dagger }a_{i\bar{\sigma}}^{\dagger}e^{i\hat{\Omega}_i}.
\label{frac2}
\end{equation}
A doped hole created by Eq. (\ref{frac2}) will generally decay into a gapped $a$-spinon and a phase shift $\hat{\Omega}$ (the holons are condensed). But the fractionalized constituents in Eq. (\ref{frac2}) can also form a bound state (i.e., a quasiparticle) at the level of RPA \cite{JHZhang2020}, which is stable and coherent within the so-called \emph{Fermi arc} regions in the Brillouin zone. Thus the present parent state may be also regarded as a pseudogap state with partially recovering of the quasiparticle excitations.

Corresponding to Eq. (\ref{frac2}), the superconducting pairing order parameter is given by  \cite{Weng2011,Ma2013}
\begin{equation}
\langle {c}_{i\uparrow } {c}_{i\downarrow }\rangle \propto \Delta_{ij}^0 \langle e^{i\left(\hat{\Omega}_i+\hat{\Omega}_j\right) } \rangle,
\label{frac3}
\end{equation}
with $\Delta_{ij}^0$ denoting the preformed pairing of the $a$-spinons in the state of Eq.(\ref{phia-0}). Then the superconducting phase coherence will be determined by $ \langle e^{i\left(\hat{\Omega}_i+\hat{\Omega}_j\right) } \rangle $, which essentially measures a vortex-antivortex pairing \cite{MW2010,Weng2011} via $\hat{\Omega}_{i,j}$ defined in Eq. (\ref{phif}). The latter in turn is determined by the $b$-spinon pairing in the state of Eq. (\ref{phirvb}) as if each $b$-spinon carries a supercurrent vortex of the holon condensate \cite{MW2010,Weng2011}. It is noted that the $a$-spinons are in s-wave pairing but the d-wave symmetry  of the hole pairing in Eq. (\ref{frac3}) is also due to the phase-string factor  
$ \langle e^{i\left(\hat{\Omega}_i+\hat{\Omega}_j\right) } \rangle $ \cite{Ma2013,Zhao2022}.

\section{Synthesis}

\subsection{Phase-string sign structure dictates a new parent state of the Mottness}

In the traditional condensed matter physics, a Fermi liquid state of many-body electrons is considered to be the \emph{parent} phase, based on which a BCS superconductivity is a spontaneous symmetry breaking or ODLRO state following the Cooper-channel instability of the Fermi liquid.  Similarly, other ODLRO states like charge-density-wave and spin-density-wave ordered phases are also the low-temperature states following, say, the Fermi surface nesting instabilities of a Fermi liquid parent state. In general, the quantum behavior of many-body electrons is within the description of the Landau paradigm in which the Fermi statistics or fermionic sign structure has laid down the essential foundation.

The validity of the Landau paradigm is ensured in a weakly correlated system in which the Fermi energy is dominant over the interaction between the electrons. However, a Landau paradigm is expected to be failed if the interaction becomes a dominant force like in the fractional quantum Hall system \cite{Laughlin1983} where the kinetic energy of the electrons gets essentially quenched by strong magnetic fields. The doped Mott insulator discussed in this article corresponds to the limit that the onsite Coulomb repulsion $U\gg t$, which represents another important case of strongly correlated electron systems that the Landau paradigm may break down.

The most striking fact in the doped Mott insulator is that the Fermi sign structure is fundamentally changed to the phase-string sign structure, which is due to the restricted Hilbert space in the presence of a Mott gap. Such a gap splits the single band of the electrons into the lower and upper Hubbard bands (cf. Fig. 1). Distinct from a trivial semiconductor, for doped holes in the lower Hubbard band or doped electrons in the upper Hubbard band, their statistical signs are no longer purely fermionic but also include exotic mutual statistics, i.e., the phase-string, which is originated from the braidings between the dopants and spins in the charge-neutral background, i.e., the filled lower Hubbard band. In general, each worldline of the dopants will be always weighed by a nontrivial phase-string sign factor (\ref{ps}) in a way precisely like the nonintegral phase factor Eq. (\ref{nonit}) due to a gauge potential \cite{Yang1974}. 

In the original RVB theory \cite{pwa,pwa1} of the doped Mott insulator, the effect of the Hilbert space restriction due to the Mott gap has been implemented by enforcing a Gutzwiller projection in the large-$U$ limit. However, the crucial phase-string sign structure, which can be rigorously identified in the $t$-$J$ and Hubbard models as the direct consequence of the Hilbert space restriction, is omitted, although a possible nontrivial response of the many-body background of the (doped) Mott insulator to each injected dopant has been conjectured \cite{pwa1,Anderson1990}. Without correctly incorporating the phase-string effect, the underlying physics would deviate from that of the true Mottness just like that a theory of the conventional metal would deviate from the Landau's Fermi liquid description should the Fermi statistics be omitted.  
 
Here the phase-string is very singular as a product of signs $\pm$, which depends on the spins exchanged with the dopant as it moves forward. A strong phase interference of different paths connecting $A$ and $B$ in Fig. \ref{psfig} is easily visualized. In particular, each spin-flip on one path can result in a total sign change or the reverse of a constructive/destructive quantum interference between the paths. As such, each dopant becomes entangled effectively with \emph{all} the neutral spins of the Mott insulator and \emph{vice versa}. So the Mottness is probably one of the most challenging non-perturbative problems of many-body quantum mechanics. Especially a long-range quantum mutual entanglement is involved with infinite degrees of freedom in the thermodynamic limit. 

Nevertheless, such an extreme strong correlation problem of the Mottness can get substantially simplified mathematically by a mutual-duality transformation. In such a dual world, the many-body electrons are fractionalized into more elementary constituents with the singular interference effect of the phase-string being exactly mapped onto three smooth topological gauge fluxes. In this new world, the matters of the fractionalized particles of infinite numbers can further experience spontaneous symmetry breaking to acquire ODLROs to gain partial rigidity, such that the internal $U(1)$ symmetries accompanying the fractionalization get suppressed (Higgsed) simultaneously. In other words, the electrons and phase-string sign structure can self-organize themselves into an emerging world of \emph{true} elementary particles that are only weakly coupled via the topological (mutual Chern-Simons) gauge fields. The validity of the fractionalization with a novel gauge structure is thus protected by the Mott gap together with the hidden ODLROs at lower temperatures. 

\subsection{A complex phase diagram with antiferromagnetism, superconductivity, and Fermi liquid as low-temperature instabilities of the parent phase}

\begin{figure}[t]
\begin{center}
\includegraphics[width=1.0\textwidth]{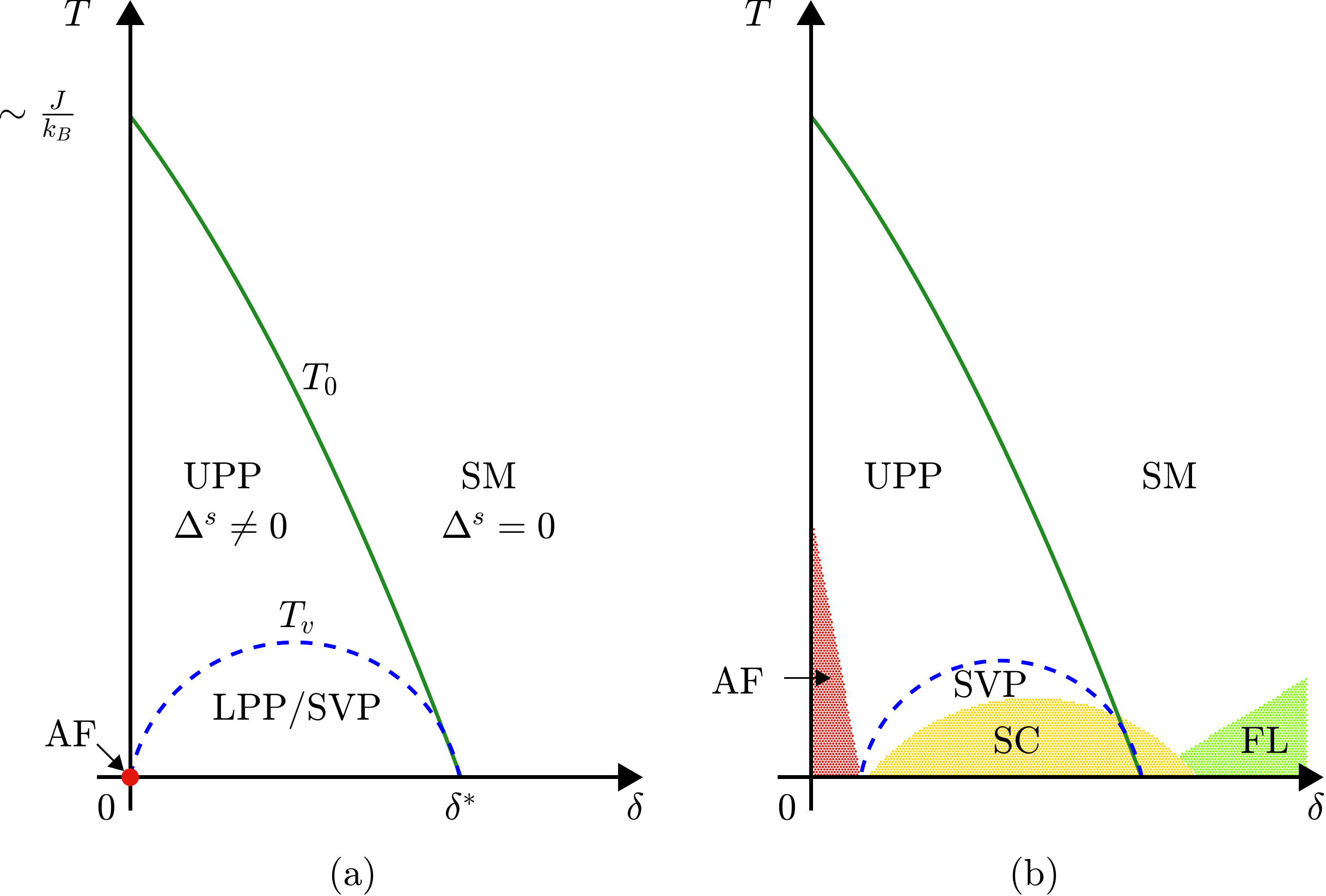}
\end{center}
\caption{(a) The phase diagram of the parent phase \cite{Weng2011,Ma2013}, in which the pseudogap phase with the bosonic RVB order parameter $\Delta^s\neq 0$ is separated into the upper (UPP) and lower (LPP) by the holon Bose-condensation at $T=T_v$. The LPP is also known as the spontaneous vortex phase (SVP); Beyond $T>T_0$, $\Delta^s= 0$ and the spin background reduces to a Curie-Weiss paramagnet with a strange-metal (SM) behavior of the dopants; (b) The antiferromagnetic (AF), superconducting (SC), and Fermi liquid (FL) as low-temperature instabilities of the parent phase in (a). }  
\label{ODLROs}
\end{figure}

The significantly reduced fermion signs due to Mottness imply the destruction of the Fermi-liquid normal state at least in the small doping concentration regime of the cuprate. Here the Fermi-liquid state is replaced by a new parent state due to the phase-string sign structure, with the multi-ODLROs characterized by a mean-field phase diagram  \cite{Ma2013} as shown in Fig. \ref{ODLROs}(a). Such a parent state characterized by Eq. (\ref{frac1}) is composed of the following essential components at $T=0$: (I) It reduces to a long-range RVB state at half filling that can highly accurately describe the AFLRO ground state of the Heisenberg model; (II) It is a d-wave superconducting state at finite doping $\delta<\delta^*$; (III) At $\delta>\delta^*$, $\Delta^s=0$ for the b-RVB order parameter of the $b$-spinons such that $|\Phi_b\rangle$ describes a Curie-Weiss paramagnetic spin background. 

In such a parent phase diagram, the $T=0$ superconducting state will be phase-disordered once $T>0^+$ by thermally excited $b$-spinons, each of which carries a supercurrent vortex formed by the holon condensate due to the mutual Chern-Simons gauge field $A^s_{ij}$. The corresponding state is called \cite{Muthu2002,Weng2006,Qi2007} the lower pseduogap phase (LPP) or spontaneous vortex phase (SVP) with a characteristic temperature scale $T_v$ of the holon Bose condensation as shown in Fig. \ref{ODLROs}(a). A true superconducting instability may further occur in the LPP/SVP at a finite transition temperature  \cite{MW2010}
\begin{equation}\label{tc}
T_c\simeq \frac{E_g}{6\mathrm{k_B}},
\end{equation}
such that the $T=0$ superconducting phase coherence can persist over to $T<T_c$ [cf. Fig. \ref{ODLROs}(b)] via making the $b$-spinons weakly ``confined'' into spinon pairs to realize the vortex-antivortex binding. In other words, in the superconducting phase, the $S=1/2$ $b$-spinons have to be bound to form the $S=0$ or $1$ resonance-like modes with a characteristic energy $E_g\neq 0$ to realize the superconducting phase coherence \cite{Ma2013,MW2010}. Here the resonance-like energy $E_g$ can be determined by the mean-field theory of the parent state, which also ensures a short-range AF correlation in the spin background self-consistently. Such a result shows that a short-range RVB state or spin liquid state can be indeed naturally realized in $|\Phi_b\rangle$ due to the phase-string effect at finite \cite{Ma2013}.  

On the other hand, by going beyond the mean-field parent state, the AFLRO in Fig. \ref{ODLROs}(a) may be restored not only at $\delta=0$, but also within a sufficiently low but finite doping regime, where $E_g\rightarrow 0$ to recover the gapless spin wave excitation \cite{Weng2007,Kou2003}. Here $E_g=0$ means the Bose-condensation of the $b$-spinons, which also implies the charge localization due to the vortex condensation associated with the spinons. It is consistent with $T_c\rightarrow 0$ according to Eq. (\ref{tc}). Namely, the AFLRO phase and superconducting phase do not overlap as controlled by the gapless and gapped spin-1 excitations, respectively. As such, how an AFLRO/insulating phase at dilute doping evolves into a short-range RVB/superconducting phase can be systematically investigated based on the parent pseudogap state by carefully considering the gauge fluctuation effects of $A^s$ and $A^h$.

At $\delta>\delta^*$, the $b$-spinon RVB order parameter $\Delta^s=0$. In such a Curie-Weiss paramagnetic spin background of $|\Phi_b\rangle$, the holons can no longer propagate coherently due to the strong destructive quantum interference of the phase-string. Only can a Landau quasiparticle effectively propagate coherently as a collective mode, which is immune from the destructive phase-string effect in this regime. A Fermi liquid behavior with full Fermi surface can be recovered in an RPA scheme beyond the parent mean-field-theory description \cite{Zhang2022}. Finally at $T>T_0$ with $\Delta^s=0$, the incoherent propagation of the holons under the frustration of the phase-string effect via the random gauge potential $A^s_{ij}$ has been shown to lead to a linear-$T$ resistivity \cite{Weng2007,Gu2005}. In other words, the strange metallic behavior observed in the cuprate may be the consequence that the two hidden ODRLOs, i.e., the $\text{b-RVB}$ order and the holon condensation in the parent state of Eq. (\ref{frac1}), are ``melted'' in the high-temperature phase of Fig. \ref{ODLROs}(a). 

\subsection{Experimental and numerical comparisons}

The cuprate superconductor is widely considered to be a class of realistic materials of strong correlation, in which the Mott gap opens up as clearly observed by experiment \cite{Wang2012}. At half-filling of the electrons, the Mott gap will render the electronic system to be an antiferromagnetic Mott insulator, and the superconductivity arises when additional charge dopants, holes or electrons, are injected into such a Mott insulator, resulting in a doped Mott insulator.  A global phase diagram including high-$T_c$ superconductivity, pseudogap, and strange metal phases has been well established experimentally in which the doping concentration and temperature seem to be two essential parameters that can be fine-tuned to result in an exceptional complexity in the phase diagram \cite{Zaanen2015}. Novel excitations revealed by the spectroscopies, thermodynamic, and transport measurements should provide direct probes of the elementary particles predicted by the Mottness. In contrast to the high-energy accelerators to test the Standard Model of the elementary particles in Nature, these experimental measurements in the high-$T_c$ cuprate have already generated a body of overly rich data in the past three decades that are capable of ``overdetermining'' the microscopic theory.

Furthermore, the finite-size exact numerics can provide another powerful and unique verification of the ground state and excitations predicted by the concrete models of the Mottness. Especially, the lattice shape, doping concentration, and the coupling parameters, are all artificially adjustable to explore a much wider phase space beyond the experimental reach to test the theoretical structure of the Mottness. In this sense, the fractionalization \emph{dual} world due to the strong correlation of the Mottness can be straightforwardly verified or falsified by both the experimental measurements and numerical experiments. 

\section{Conclusion}

The Mott physics provides a striking example to illustrate ``more is different'' or the emergent phenomenon in many-body quantum mechanics \cite{pwa1972,LP2000}. A single band of electrons in the presence of the onsite Coulomb repulsion would behave generally as a Fermi liquid \cite{AGD1963,Pines1965} in the limit that the Fermi energy is dominant over the interaction. But in the opposite limit of $t\ll U$, a Mott gap opens up and the low-energy physics inside the gap will be completely changed. Here the Fermi statistics is replaced by the phase-string sign structure as the new organizing principle. The latter is equivalent to a mutual semionic statistics between the fractionalized coordinates in the restricted Hilbert space of the Mottness. In this brief review, a consistent picture based on a systematic study has been presented, in which the key emphasis is on that the new elementary particles with weakly-coupled mutual Chern-Simons gauge fields are the faithful representation of the Mottness in a dual world. In particular, near the half-filling, these quantum constituents form multiple hidden ODLROs in the thermodynamic limit. It is effectively a rigid system with low-energy gaps or short-range orders, but is short of having a true ODLRO at $T>0^+$. It resembles a pseudogap state with a gapless quasiparticles emerging in portions (Fermi arcs) of the Brillouin zone as a collective mode of the elementary particles. Antiferromagnetically ordered phase, superconducting phase, and the Femi liquid phase, may be regarded as the low-temperature instabilities of such a parent state at different doping regimes. As such, a series of sharp and unique predictions can be explicitly made for the high-$T_c$ cuprate based on the Mottness.

\section*{Acknowledgments}

Stimulating discussions with Shuai Chen, Donna Sheng, Jan Zaanen, and Jia-Xin Zhang are acknowledged. A partial support of this work by the MOST of China grant No. 2017YFA0302902 is also acknowledged.


\begin{thebibliography}{99}

\bibitem{Yang1974} C. N. Yang, Phys. Rev. Lett. \textbf{33}, 445 (1974); T. T. Wu and C. N. Yang, Phys. Rev. D \textbf{12}, 3845 (1975).


\bibitem{AGD1963} A. A. Abrikosov, L. P. Gor'kov, and E. Dzyaloshinskii, \emph{Method of Quantm Field Theory in Statistical Physics }, (Prentice-Hall, Englewood Cliffs, 1963). 

\bibitem{Pines1965} D. Pines and P. Nozieres, \emph{Theory of Quantum Liquids }, (Benjamin, New York 1965). 

\bibitem{LNW2006} For a review, see, P. A. Lee, N. Nagaosa, and X. G. Wen,
Rev. Mod. Phys. \textbf{78}, 17 (2006).

\bibitem{pwa}  P. W. Anderson, Science \textbf{235}, 1196 (1987).

\bibitem{pwa1}  P. W. Anderson, \emph{The Theory of
Superconductivity in the High $T_c$ Cuprates}, (Princeton Univ. Press,
Princeton, 1997).

\bibitem{Wang2012} C. Ye, P. Cai, R. Z. Yu, X. D. Zhou, W. Ruan, Q. Q. Liu, C. Q. Jin, and Y. Y. Wang, Nat. Commun. 4, 1365 (2013).

\bibitem{Zaanen2015} B. Keimer, S. A. Kivelson, M. R. Norman, S. Uchida, and J. Zaanen, Nature \textbf{518}, 179 (2015).

\bibitem{Weng2007} For an earlier review, see, Z. Y. Weng, Intl. J. Mod. Phys. B \textbf{21}, 773 (2007); arXiv: 0704.2875.

\bibitem{Weng2011r} Z. Y. Weng, Front. Phys. \textbf{6}, 370 (2011); arXiv: 1110.0546.

\bibitem{Zaanen2009} J. Zaanen and B. J. Overbosch, Phil. Trans. R. Soc. A 
\textbf{369}, 1599 (2011); arXiv:0911.4070.

\bibitem{pwa2} P. W. Anderson, P. A. Lee, M. Randeria, T. M. Rice, N.
Trivedi, and F. C. Zhang, J. Phys.: Condens. Matter \textbf{16}, R755
(2004), and references therein.

\bibitem{gros07} For a review, see, B. Edegger, V.N. Muthukumar, and C.
Gros, Adv. Phys., \textbf{56}, 927 (2007).

\bibitem{Wen2019} X.G. Wen, arXiv:1906.05983.  

\bibitem{Weng1996} D. N. Sheng, Y. C. Chen, and Z. Y. Weng, Phys. Rev. Lett. 
\textbf{77}, 5102 (1996).

\bibitem{WWZ2008} K. Wu, Z. Y. Weng, and J. Zaanen, Phys. Rev. B \textbf{77},
155102 (2008).

\bibitem{Wu1984} Y. S. Wu, Phys. Rev. Lett. \textbf{52}, 2103 (1984).

\bibitem{Zhang2014} L. Zhang and Z. Y. Weng, Phys. Rev. B \textbf{90}, 165120 (2014).  

\bibitem{Anderson1990} P. W. Anderson, Phys. Rev. Lett. \textbf{64}, 1839 (1990).

\bibitem{Anderson1967a} P. W. Anderson, Phys. Rev. Lett. \textbf{18}, 1049 (1967).

\bibitem{Anderson1967b} P. W. Anderson, Phys. Rev. \textbf{164}, 352 (1967).

\bibitem{Weng1997} Z. Y. Weng, D. N. Sheng, Y.-Chen, and C. S. Ting, Phys.
Rev. B \textbf{55}, 3894 (1997).

\bibitem{Weng1998} Z. Y. Weng, D. N. Sheng, and C. S. Ting, Phys.
Rev. Lett. \textbf{80}, 5401 (1998).

\bibitem{Weng2011} Z. Y. Weng, New J. Phys, \textbf{13}, 103039 (2011).

\bibitem{Zhu2013} Z. Zhu, H.-C. Jiang, Y. Qi, C. Tian, Z.-Y. Weng, Sci. Rep. \textbf{3}, 2586 (2013).

\bibitem{Wilczek1990} F. Wilczek, \emph{Fractional Statistics and Anyon Superconductivity}, (World Scientific Publishing Co. Pte. Ltd. 1990). 

\bibitem{Marshall} W. Marshall, Proc. R. Soc. A\textbf{232}, 48 (1955).

\bibitem{Zheng2018} W. Zheng, Z. Zhu, D.N. Sheng, and Z.Y. Weng, Phys. Rev. B \textbf{98},165102 (2018).

\bibitem{Zhu2016} Z. Zhu, Q.-R. Wang, D.N. Sheng, and Z.-Y. Weng, Nucl. Phys. B \textbf{903}, 51 (2016).

\bibitem{Zheng2018r} W. Zheng and Z.-Y. Weng, Sci. Rep. \textbf{8}, 3612 (2018).

\bibitem{Zhu2015} Z. Zhu and Z. Y. Weng, Phys. Rev. B \textbf{92}, 235156 (2015).

\bibitem{Zhu2018} Z. Zhu, D. N. Sheng, and Z. Y. Weng, Phys. Rev. B \textbf{98}, 035129 (2018).

\bibitem{Sun2019} R.-Y. Sun, Z. Zhu, and Z. Y. Weng, Phys. Rev. Lett. \textbf{123}, 016601 (2019).

\bibitem{He2016} R.-Q. He and Z.-Y. Weng, Sci. Rep. \textbf{6}, 35208 (2016).

\bibitem{Ho2016} Z. Zhu, Z.-Y. Weng, and T.-L. Ho, Phys. Rev. A \textbf{93}, 033614 (2016).

\bibitem{Tohyama2021} Kazuya Shinjo, Shigetoshi Sota, Takami Tohyama, Phys. Rev. B \textbf{103}, 035141 (2021).


\bibitem{Zhu2014} Z. Zhu, H.C. Jiang, D.N. Sheng, and Z.-Y. Weng, Sci. Rep. \textbf{4}, 5419 (2014).

\bibitem{Zhu2018pairing} Z. Zhu, D. N. Sheng, and Z.-Y. Weng, Phys. Rev. B \textbf{97},115144 (2018).
        
\bibitem{Jiang2020} Hong-Chen Jiang, Shuai Chen, and Zheng-Yu Weng,  Phys. Rev. B \textbf{102}, 104512 (2020).

\bibitem{Sun2020} R.-Y. Sun, Z. Zhu, and Z. Y. Weng, Phys. Rev. Res. \textbf{2}, 033007 (2020).

\bibitem{Jiang2018} Hong-Chen Jiang, Zheng-Yu Weng, and Steven A. Kivelson, Phys. Rev. B \textbf{98}, 140505 (2018).

\bibitem{Jiang2019} Y. F. Jiang, J. Zaanen, T. P. Devereaux, H. C. Jiang, Phys. Rev. Res. \textbf{2}, 033073 (2020).

\bibitem{White2022} S. Jiang, D. J. Scalapino, S. R. White, PNAS \textbf{118}, 2109978118 (2021).

\bibitem{ZhangJX2022} J.-X. Zhang, \emph{et al.},  in preparation.

\bibitem{Ma2013} Y. Ma, P. Ye, and Z. -Y. Weng, New J. Phys, \textbf{16}, 083039 (2014).
 
\bibitem{Liang1988} S. Liang, B. Doucot, P. W. Anderson, Phys. Rev. Lett. \textbf{61}, 365 (1988).

\bibitem{Wang2015}  Q.-R. Wang,  Z. Zhu, Y. Qi, Yang, Z.Y. Weng, arXiv1509.01260.

\bibitem{Chen2019} S. Chen, Q.R. Wang, Y. Qi, D.N. Sheng, and Z. Y. Weng, Phys. Rev. B \textbf{99}, 205128 (2019).

\bibitem{Chen2022} S. Chen, Z.-Y. Weng, and J. Zaanen, Phys. Rev. B \textbf{105}, 075136 (2022).

\bibitem{Chen2018} S. Chen, Z. Zhu, and Z.-Y. Weng, Phys. Rev. B \textbf{98}, 245138 (2018).

\bibitem{Zhao2022} J.-Y. Zhao, S. Chen, H.-K. Zhang, and Z.-Y. Weng, Phys. Rev. X \textbf{12}, 011062 (2022).

\bibitem{note1} Note that here two twisted holes are defined by using the same sign of the phase-string factor in Eq. (\ref{eqn:ctilde}), a topological correction to the half-filling ground state should be included in $|\text{b-RVB} \rangle$ according to Ref. \onlinecite{Zhao2022}.    

\bibitem{Yang-ODLRO} C.N. Yang, Rev. Mod. Phys. \textbf{34}, 694 (1962).

\bibitem{KQW2005} S. P. Kou, X. -L. Qi, and Z. -Y. Weng,
Phys. Rev. B \textbf{71}, 235102 (2005).

\bibitem{Ye2011} P. Ye, C. Tian, X. -L. Qi, and Z.-Y. Weng, Phys. Rev.
Lett. \textbf{106}, 147002 (2011); Nucl. Phys. B \textbf{854}, 815 (2012).

\bibitem{JHZhang2020} J.-H. Zhang, S. Li, Y. Ma, Y. Zhong, H. Ding, and Z. -Y. Weng, Phys. Rev. Res. \textbf{2}, 023398 (2020).


\bibitem{MW2010} J. W. Mei and Z. Y. Weng, Phys. Rev. B \textbf{81}, 014507
(2010).

\bibitem{Laughlin1983} R.B. Laughlin, Phys. Rev. Lett. \textbf {50}, 1395 (1983).

\bibitem{Muthu2002} V.N. Muthukumar and Z.Y. Weng, Phys. Rev. B \textbf{65}, 174511
(2002).


\bibitem{Weng2006} Z. Y. Weng and X. L. Qi, Phys. Rev. B \textbf{74}, 144518 (2006).

\bibitem{Qi2007} X. L. Qi and Z. Y. Weng, Phys. Rev. B \textbf{76}, 104502 (2007).

\bibitem{Kou2003} S.-P. Kou and Z.-Y. Weng, Phys. Rev. Lett. \textbf{90}, 157003
(2003).

\bibitem{Zhang2022} J.-X. Zhang, \emph{et al. } in preparation (2022).

\bibitem{Gu2005} Z.-C. Gu and Z.-Y. Weng, Phys. Rev. B \textbf{72}, 104520 (2005).

\bibitem{pwa1972} P. W. Anderson, Science, \textbf{177}, 393 (1972). 

\bibitem{LP2000} R. B. Laughlin and D. Pines, PNAS, \textbf{97}, 28 (2000).



    
\end{thebibliography}
\end{document}